\documentclass[12pt]{iopart}

\usepackage{graphics}
\usepackage{color}
\usepackage{bm}

\def\PRA{\textit{Phys.~Rev.~A} }

\def\JPB{\textit{J.~Phys.~B: At. Mol. Opt. Phys.} }
\def\PRL{\textit{Phys.~Rev.~Lett.} }

\newcommand{\myscaleboxa}[1]{\scalebox{0.33}[0.33]{#1}}
\newcommand{\myscaleboxb}[1]{\scalebox{0.42}[0.46]{#1}}

\usepackage{epsfig}

\usepackage{hyperref}

\begin{document}

\title{Calibration of distorted wave Born approximation for electron impact
excitation of Ne and Ar at incident energies below 100 eV}

\author{Yaqiu Liang$^1$, Zhangjin Chen$^2$, D. H. Madison$^3$ and C. D. Lin$^2$}

\address{$^1$ College of Physics, Liaoning University, Shenyang 110036, People's
Republic of China}

\address{$^2$ J. R. Macdonald Laboratory, Physics Department, Kansas
State University, Manhattan, Kansas 66506-2604, USA}

\address{$^3$ Physics Department, Missouri University of Science and Technology,
Rolla, Missouri 65401, USA}

\begin{abstract}

We calibrate the distorted wave Born approximation (DWBA) for
electron impact excitation processes empirically. Differential cross
sections (DCS) for the excitation of the $2p^53s$, $2p^53p$,
$2p^54s$, and $2p^54p$ configurations of Ne and the $3p^54s$ and
$3p^54p$ configurations of Ar by electron impact are calculated
using DWBA for incident energies between 20 and 100 eV. The
calculated results are compared with the absolute experimental
measurements and other theoretical results. We found that the
structure of the DCS can be well reproduced by the DWBA model while
the magnitude is overestimated for most cases considered here. The
differences in magnitude between DWBA and experiment are used to
test the calibration of DWBA such that the DWBA can be used to
describe laser-induced electron impact excitation processes. These
processes are involved in the non-sequential double ionization of
atoms in strong laser fields.

\end{abstract}

\pacs{34.80.Dp, 34.50.Rk}

\submitto{\JPB}

\maketitle

\section{Introduction}

The process of electron impact excitation of atoms and ions is one
of the most basic and important processes in atomic physics.
Numerous theoretical methods have been used for the process
calculations, including distorted wave Born approximation
(DWBA)~\cite{Madison-jpb1987}, second-order distorted wave
model~\cite{Madison-jpb1983}, $R$-matrix
method~\cite{Bartschat-jpb1997}, and convergent close-coupling (CCC)
calculations~\cite{Igor-pra-1992}, among which the DWBA is the
simplest. The sophisticated theoretical models, such as CCC and
$R$-matrix method, are capable of reproducing accurate angular
differential cross sections (DCS), as well as the absolute
magnitude. They are more suitable for low incident energies. For
higher energies, both the integrated and differential cross sections
predicted by DWBA are  fairly accurate. However, it has been well
recognized that, at low energies, the total cross sections (TCS)
predicted by the DWBA signicantly exceed the experimental values.
Ideally, one would use the $R$-matrix approach for low energies, the
DWBA for high energies, but at intermediate energies, neither method
is efficient if a large amount of data is needed.

The purpose of this work is to correct (by renormalization) the DWBA
predictions empirically such that the DCS calculated from DWBA can
be used for low collision energies. We will use the empirical
formula proposed by Tong \emph{et al}.~\cite{Tong03pra} for the
total excitation cross sections. Our ultimate objective is to apply
the calibrated DWBA (C-DWBA) to simulate the correlated momentum
distributions in nonsequential double ionization (NSDI) of atoms in
strong laser fields.

The process of NSDI of atoms in linearly polarized laser pulses is
one of the most interesting and challenging topics in strong field
physics. In NSDI, one electron that is first released near the
maximum of the oscillating electric field may be driven back to
revisit the parent ion when the electric field is near zero. When
the returning electron collides with the parent ion with energies
above the ionization threshold, it may kick out another bound
electron, resulting in an (e,~2e)-like process. The returning
electron may also excite the bound electron to a higher excited
state which is subsequently tunnel ionized when the electric field
increases again. Since the year of 2000, complete experimental
measurements on the full momentum vectors of the two outgoing
electrons along the direction of polarization of the laser pulse
have become available~\cite{Weber_Nature,Staudte_PRL07}, and a
number of theoretical studies have also been carried out.

Recently, Chen \emph{et al}.~\cite{Chen_pra09} have developed a
quantitative rescattering (QRS) theory which has been applied to
various  rescattering processes induced by short intense laser
pulses~\cite{Lin-jpb-review,Chen-pra09R,Liang-pra10,Sam-NSDI,Chen-prl10,Chen-pra10}.
The significant advantage of the QRS theory is that it treats the
rescattering processes in the laser field as laser-\emph{free}
scattering processes, where the laser-induced returning electrons
are described by a wavepacket. The QRS enables us to simulate
two-dimensional correlated momentum distributions for NSDI
\emph{quantitatively} by calculating the triple differential cross
sections (TDCS) for (e,~2e)~\cite{Chen-prl10} and the DCS for
electron impact excitation of ions~\cite{Chen-pra10}. However, to
obtain the correlated momentum spectra that can be compared with
experimental measurements, one needs to evaluate the TDCS for
(e,~2e) and the DCS for excitation for all possible momenta of the
returning electrons. For NSDI of atoms in strong laser pulses, the
highest energy, $E_i^{\mathrm {max}}$, of the returning (incident)
electron  is determined by the laser field, which is less than 100
eV for typical 800 nm lasers. To simulate the correlated momentum
distributions for NSDI, the DCS's for electron impact excitation of
the parent ion at all incident energies from threshold to
$E_i^{\mathrm {max}}$ are needed. To reduce the computational effort
a simple and efficient theoretical model is desirable. In this work,
we develop the C-DWBA for this purpose.

The organization of this paper is as follows: In section II, the
basic theory of DWBA for electron impact excitation is presented and
the method to calibrate DWBA is introduced. In section III, the DCS
of DWBA for electron impact excitation of Ne and Ar at incident
energies below 100 eV are normalized and compared with the absolute
experimental data. The normalization factors are then used to test
the calibration for DWBA.

Atomic units are used in this paper unless otherwise specified.

\section{Theory}

In this section, we present the DWBA theory on electron impact
excitation of atoms and the method to calibrate the DWBA at low
energies. The formulas presented here are generic and therefore can
be easily applied to the processes of electron impact excitation of
ions which are involved in NSDI.

\subsection{DWBA}

Suppose we have an electron with momentum $\mathbf{k}_i$ which
collides with an atom $A$, after the collision, the scattered
electron has momentum $\mathbf{k}_f$, and one bound electron in atom
$A$ is excited to a higher energy bound state. In the frozen core
approximation, the ``exact" Hamiltonian for the whole system is
\begin{eqnarray}
H=-\frac{1}{2}\nabla_1^2+V_{A^+}(r_1)-\frac{1}{2}\nabla_2^2+V_{A^+}(r_2)+\frac{1}{r_{12}}.
\end{eqnarray}
where $\mathbf{r}_1$ and $\mathbf{r}_2$ are the position vectors for
the projectile and the bound state electron with respect to the
nucleus, respectively. This Hamiltonian can be rewritten
approximately as
\begin{eqnarray}
H_j=-\frac{1}{2}\nabla_1^2+U_j(r_1)-\frac{1}{2}\nabla_2^2+V_{A^+}(r_2)
\ \ \ \ \ (j=i,f).
\end{eqnarray}
In this equation, $U_i$ ($U_f$) is the distorting potential used to
calculate the initial (final) state wave function
$\chi_{\mathbf{k}_i}$ ($\chi_{\mathbf{k}_f}$) for the projectile. In
the distorted wave Born approximation, the direct transition
amplitude for excitation from an initial state $\Psi_i$ to a final
state $\Psi_f$ is expressed by
\begin{eqnarray}
\label{t-matrix}f=
\langle\chi^-_{\mathbf{k}_f}(1)\Psi_f(2)|V_i|\Psi_i(2)\chi^+_{\mathbf{k}_i}(1)\rangle,
\end{eqnarray}
where $V_i$ is the perturbation interaction,
\begin{eqnarray}
\label{pot}V_i=H-H_i=\frac{1}{r_{12}}+V_{A^+}(r_1)-U_i(r_1).
\end{eqnarray}
In Eq.~(\ref{t-matrix}), the initial and final state wave functions
for the projectile satisfy the differential equation
\begin{eqnarray}
\label{prjectile}\left[-\frac{1}{2}\nabla_1^2+U_j(r_1)-\frac{1}{2}k^2_j\right]\chi_{\mathbf{k}_j}(\mathbf{r}_1)=0
\ \ (j=i,f),
\end{eqnarray}
and the bound state wave functions are eigenfunctions of the
equation
\begin{eqnarray}
\label{initial2}\left[-\frac{1}{2}\nabla_2^2+V_{A^+}(r_2)-\epsilon_j\right]\Psi_j(\mathbf{r}_2)=0
\ \ (j=i,f),
\end{eqnarray}
where $\epsilon_j$ ($j=i,f$) are the corresponding eigenenergies of
the initial and final bound states which can be expressed as
\begin{eqnarray}
\Psi_j(\textbf{r})=\psi_{N_jLj}(r)Y_{L_jM_j}(\hat{\textbf{r}}) \ \ \
(j=i,f).
\end{eqnarray}

The exchange scattering amplitude is given by
\begin{eqnarray}
\label{exchange-T-martrix}
g=\langle\Psi_f(1)\chi^-_{\mathbf{k}_f}(2)|V_i|\Psi_i(2)\chi^+_{\mathbf{k}_i}(1)\rangle.
\end{eqnarray}

Finally, the differential cross section for electron impact
excitation is given by
\begin{eqnarray}
\label{dcs} \frac{d\sigma}{d\Omega}&=&N(2\pi)^4
\frac{k_f}{k_i}\frac{1}{2L_i+1}\nonumber\\
&\times& \sum^{+L_i}_{M_i=-L_i}\sum^{+L_f}_{M_f=-L_f}
\left(\frac{3}{4}|f-g|^2 +\frac{1}{4}|f+g|^2\right).
\end{eqnarray}
The prefactor $N$ in Eq.~(\ref{dcs}) denotes the number of electrons
in the subshell from which one electron is excited.

The distorting potentials, $U_i$ and $U_f$, used in
Eq.~(\ref{prjectile}) to calculate the wave functions for the
projectile in the initial and final states, respectively, are not
determined directly by the formalism. Here, we use static potentials
which take the form as
\begin{eqnarray}
\label{dist-pot}U_j(r_1)=V_{A^+}(r_1)+\int
d\mathbf{r}_2\frac{|\Psi_j(\mathbf{r}_2)|^2}{r_{12}} \ \ \ \
(j=i,f).
\end{eqnarray}
As shown previously, $V_{A^+}(r)$ in Eq.~(\ref{dist-pot}) is the
atomic potential used to evaluate eigenstate wave functions of the
bound state electron. Here we use the effective potential from Tong
and Lin~\cite{Tong} based on single active electron approximation,
which is given by
\begin{eqnarray}
V_{A^+}(r)=-\frac{1+a_1e^{-a_2r}+a_3re^{-a_4r}+a_5e^{-a_6r}}{r},
\end{eqnarray}
where the parameters $a_i$, as given explicitly in table 1 in Tong
and Lin~\cite{Tong}, are obtained by fitting the calculated binding
energies from this potential to the experimental ones of the ground
state and the first few excited states of the target atom.

\subsection{Calibration of DWBA}

The overestimate of DWBA on DCS can be corrected by using the
empirical method proposed by Tong \emph{et al}.~\cite{Tong03pra} to
evaluate the total cross sections for electron impact excitation:
\begin{eqnarray}
\label{xm1}\sigma_{\mathrm {Tong}}(E_i)=\alpha\frac{\pi}{\Delta
E^2}e^{1.5(\Delta E-\epsilon)/E_i}f\left(\frac{E_i}{\Delta
E}\right),
\end{eqnarray}
where
\begin{eqnarray}
\label{xm2}f(x)=\frac{1}{x}\left[\beta \ln x
-\gamma\left(1-\frac{1}{x}\right)+\delta \frac{\ln x}{x} \right].
\end{eqnarray}
In Eq.~(\ref{xm1}), $\Delta E$ is the excitation energy for a given
transition, $\epsilon$ is the eigenenergy of the corresponding
excited state. The original formula given by Tong \emph{et al}. does
not have the prefactor $\alpha$ which is added in the present work
to ensure that it reproduces the same cross sections as those from
DWBA at high energies. The parameters in Eq.~(\ref{xm2}) have been
obtained initially by fitting to the convergent-close coupling
excitation cross sections for hydrogen and He$^+$. Explicitly, these
parameters are $\beta=0.7638$, $\gamma=1.1759$, and $\delta=0.6706$.

The total cross section of DWBA at fixed incident energy
$E_i=k_i^2/2$ can be obtained from Eq.~(\ref{dcs}) by
\begin{eqnarray}
\label{total-dwba}\sigma_{\mathrm {DWBA}}(E_i)=\int \frac{d
\sigma}{d \Omega} d \hat{\bf k}_f.
\end{eqnarray}
By matching the total cross sections from Eq.~(\ref{xm1}) with those
from Eq.~(\ref{total-dwba}) at high incident energies, say $E_i=500$
eV, one obtains the prefactor $\alpha$ in Eq.~(\ref{xm1}) for
excitation of each configuration. To calibrate the DWBA at low
energies, we define a scaling factor
\begin{eqnarray}
\label{xm3}\mathcal{C}(E_i)=\sigma_{\mathrm
{Tong}}(E_i)/\sigma_{\mathrm {DWBA}}(E_i).
\end{eqnarray}
It is the the scaling factor in Eq.~({\ref{xm3}) that should be used
to normalize the differential cross sections of DWBA at different
incident energies.

\section{Results and discussions}

The perturbative nature of DWBA makes it overestimate electron
impact excitation cross sections of ions at low energies. To
calibrate the DWBA theory, one should compare its predictions to
accurate theoretical results or absolute experimental measurements.
Unfortunately, neither are easily available for atomic and molecular
ions. Thus we use neutral Ne and Ar atoms for the calibration.

\begin{figure}
\centering \mbox{\rotatebox{270}{\myscaleboxa{
\includegraphics{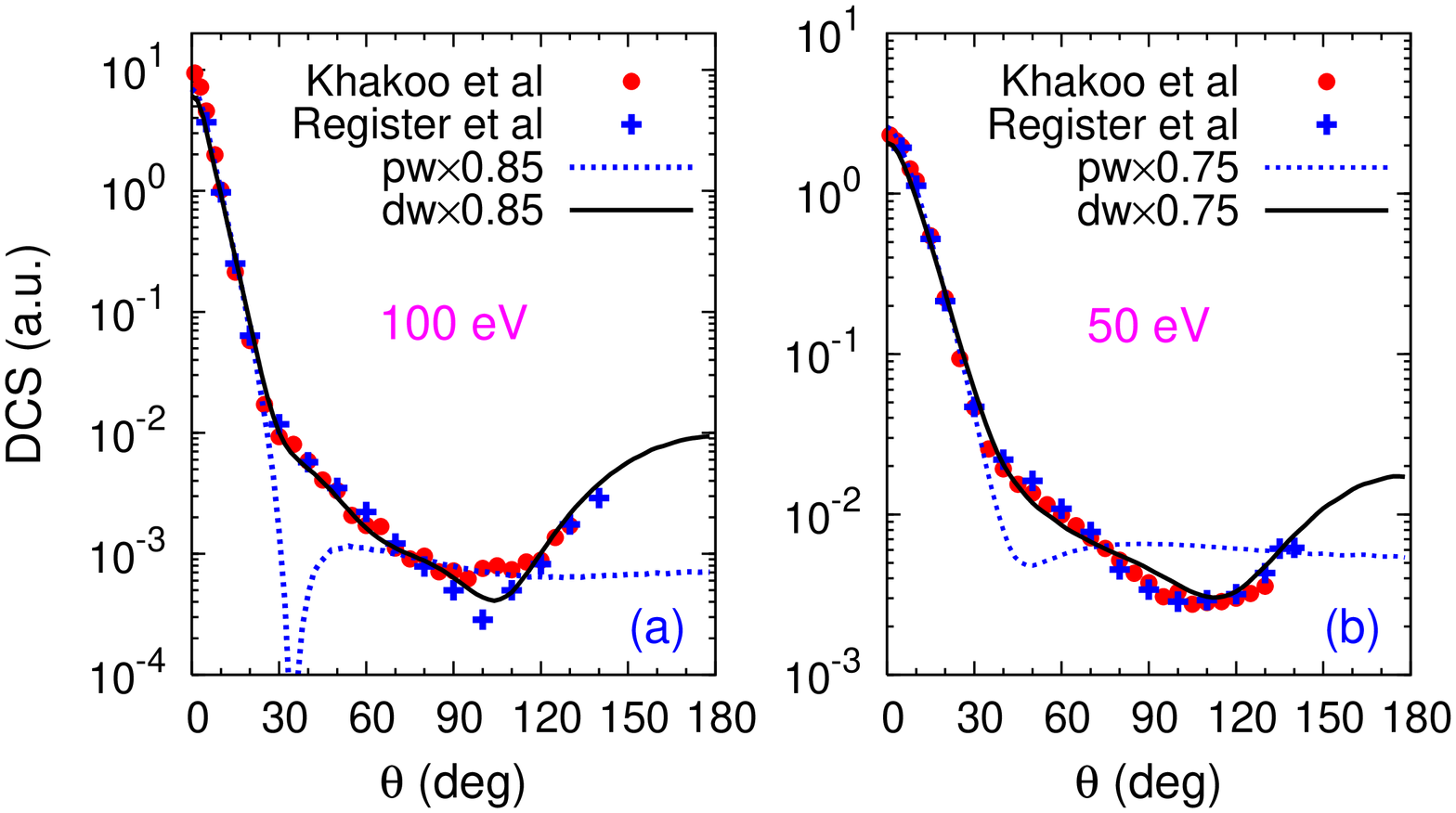}}}}
\mbox{\rotatebox{270}{\myscaleboxa{
\includegraphics{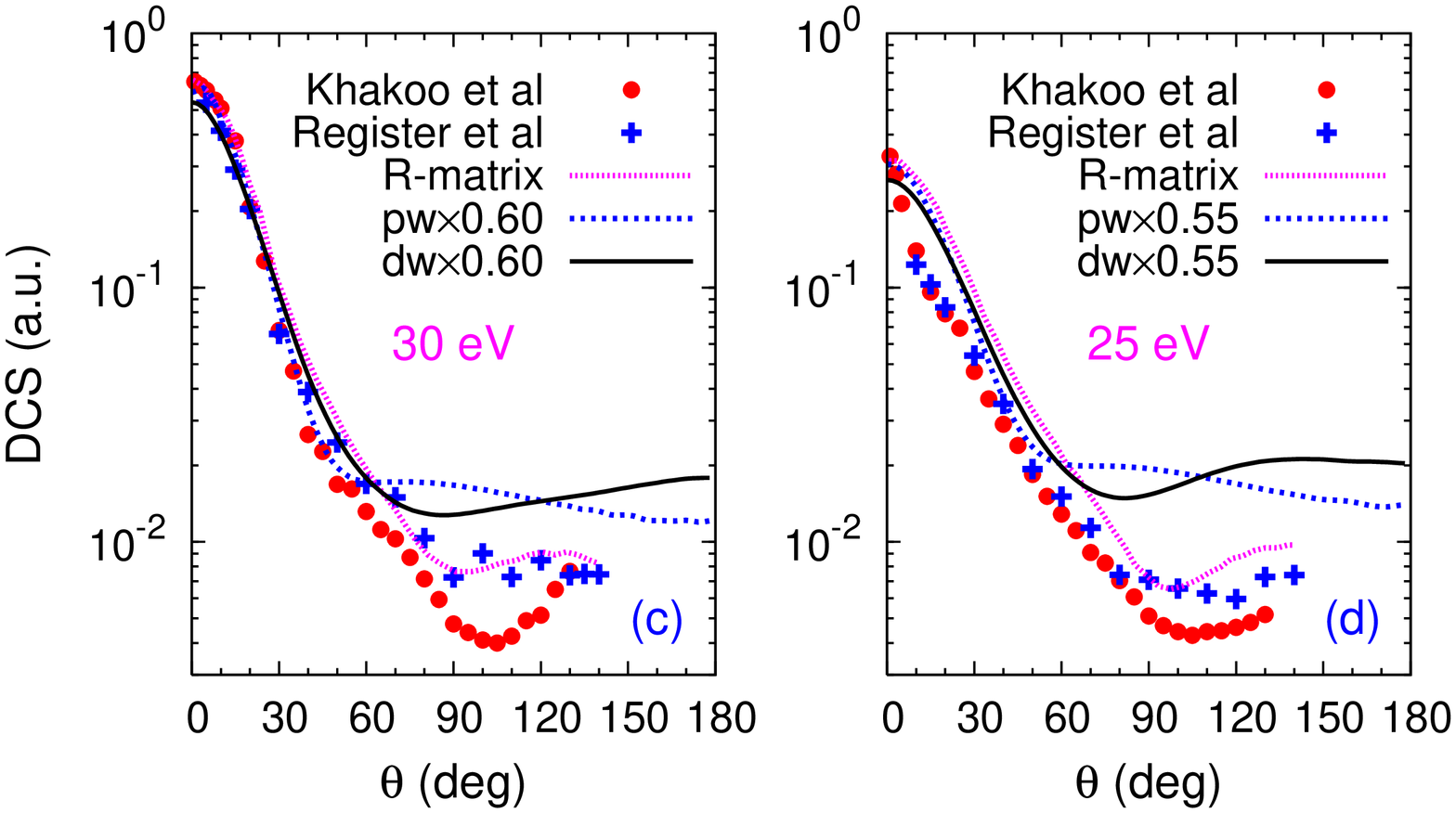}}}}
\caption{DCS for the excitation of the $2p^53s$ configuration of Ne
by electron impact at incident energies of (a) 100 eV, (b) 50 eV,
(c) 30 eV, and (d) 25 eV. The absolute experimental measurements are
from Register \emph{et al}.~\cite{Register} and Khakoo \emph{et
al}.~\cite{Khakoo}. For incident energies of 30 and 25 eV, the
results of $R$-Matrix from Khakoo \emph{et al}.~\cite{Khakoo} are
also plotted for comparison.}
\end{figure}

\begin{figure}
\centering \mbox{\rotatebox{270}{\myscaleboxa{
\includegraphics{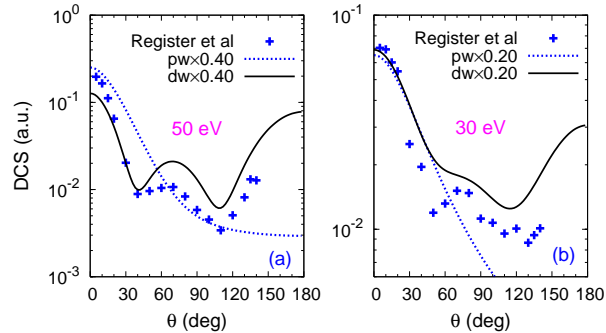}}}}
\caption{Same as Fig.~1 but for configuration of $2p^53p$ at
incident energies of (a) 50 eV and (b) 30 eV.}
\end{figure}

\begin{figure}
\centering \mbox{\rotatebox{270}{\myscaleboxa{
\includegraphics{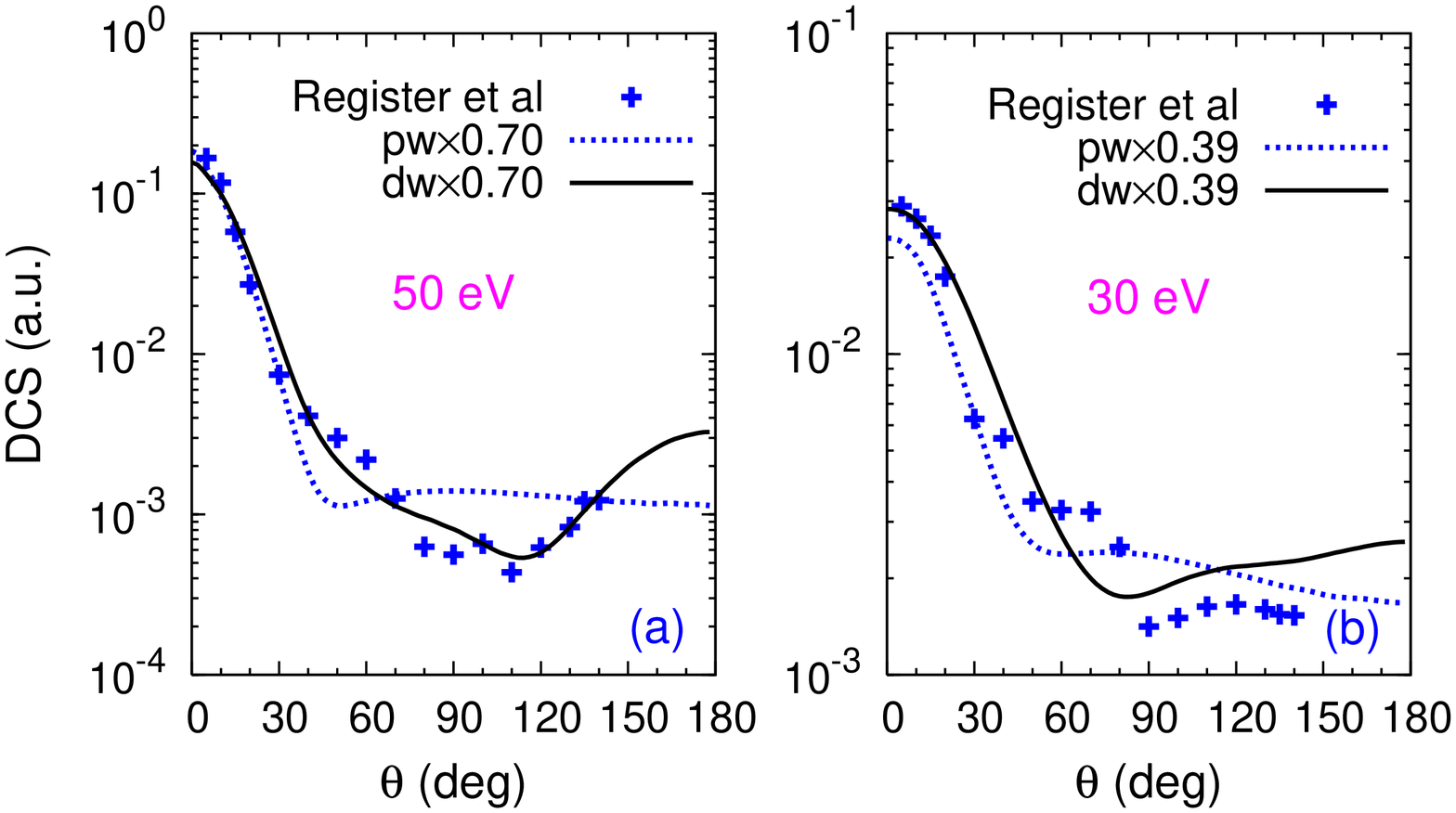}}}}
\caption{Same as Fig.~1 but for the $2p^54s$ configuration at
incident energies of (a) 50 eV and (b) 30 eV.}
\end{figure}

\begin{figure}
\centering \mbox{\rotatebox{270}{\myscaleboxa{
\includegraphics{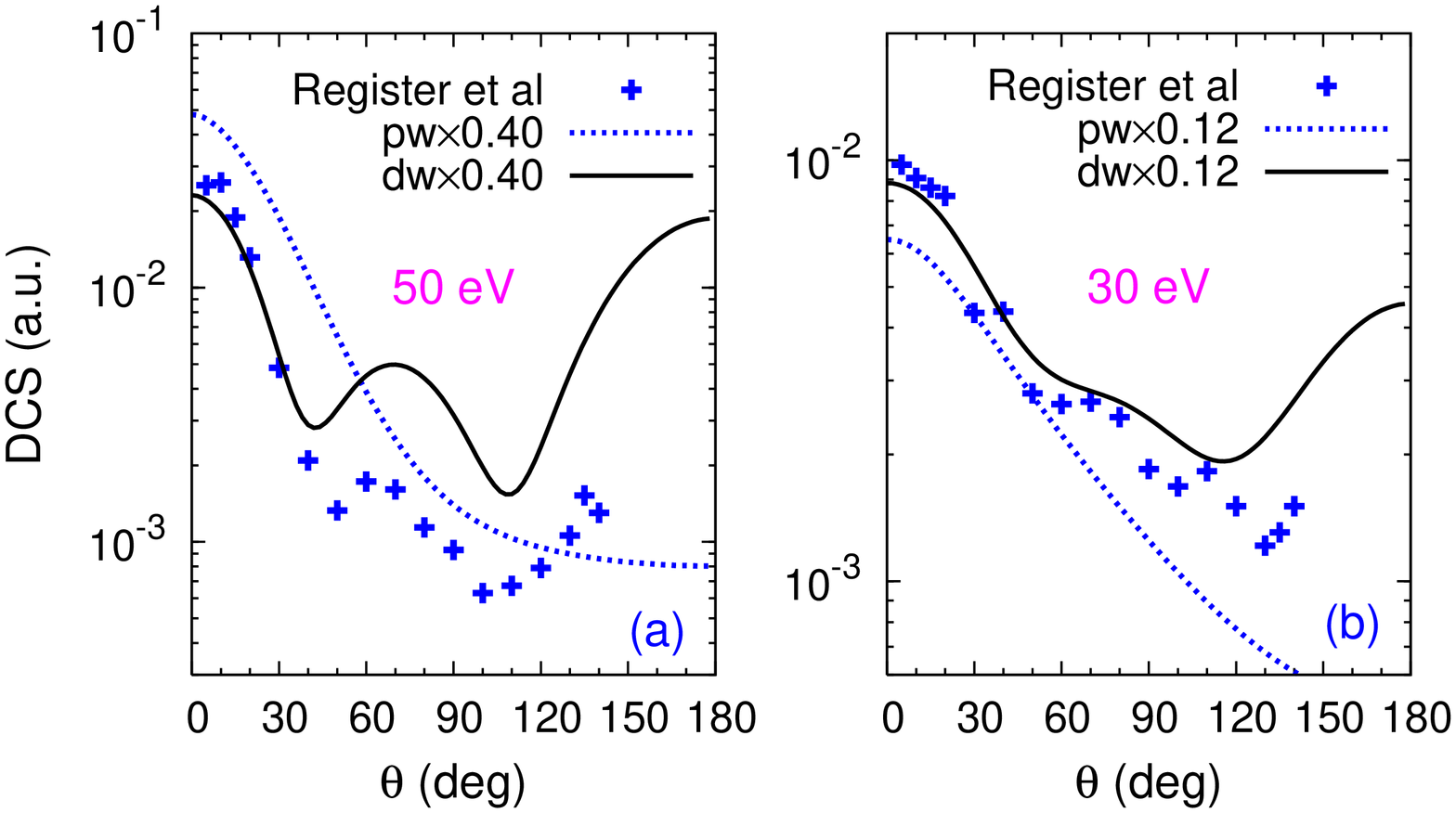}}}}
\caption{Same as Fig.~1 but for the $2p^54p$ configuration at
incident energies of (a) 50 eV and (b) 30 eV.}
\end{figure}

In Figures~1-4, the DCS's for the excitation of the $2p^53s$,
$2p^53p$, $2p^54s$ and $2p^54p$ configurations of Ne for incident
energies below 100 eV from DWBA are compared with the absolute
experimental data~\cite{Register,Khakoo}. For the $2p^53s$
configuration, the DCS's from the $R$-matrix theory are also plotted
for incident energies of 30 and 25 eV. It can be seen from Figs.~1-4
that the DWBA overestimates the DCS's for all the cases considered
here. To get best overall agreement, different normalization factors
are assigned to the DWBA for different configurations and different
incident energies. For incident energies below 30 eV, the DCS's of
DWBA could be 2-7 times higher than the experimental measurements.
To see the distorting effect from the DWBA, the results of plane
wave Born approximation (PWBA), in which plane waves are used to
describe the projectile electron in both the initial and final
states, are also displayed for comparison. For scattering angles
greater than 30$^{\circ}$, one can see that the PWBA fails
completely in predicting the angular distributions even for incident
energy of 100 eV. In contrast, the enhanced DCS's for backward
scattering observed in experiment are well reproduced by the DWBA.

\begin{figure}
\centering \mbox{\rotatebox{270}{\myscaleboxa{
\includegraphics{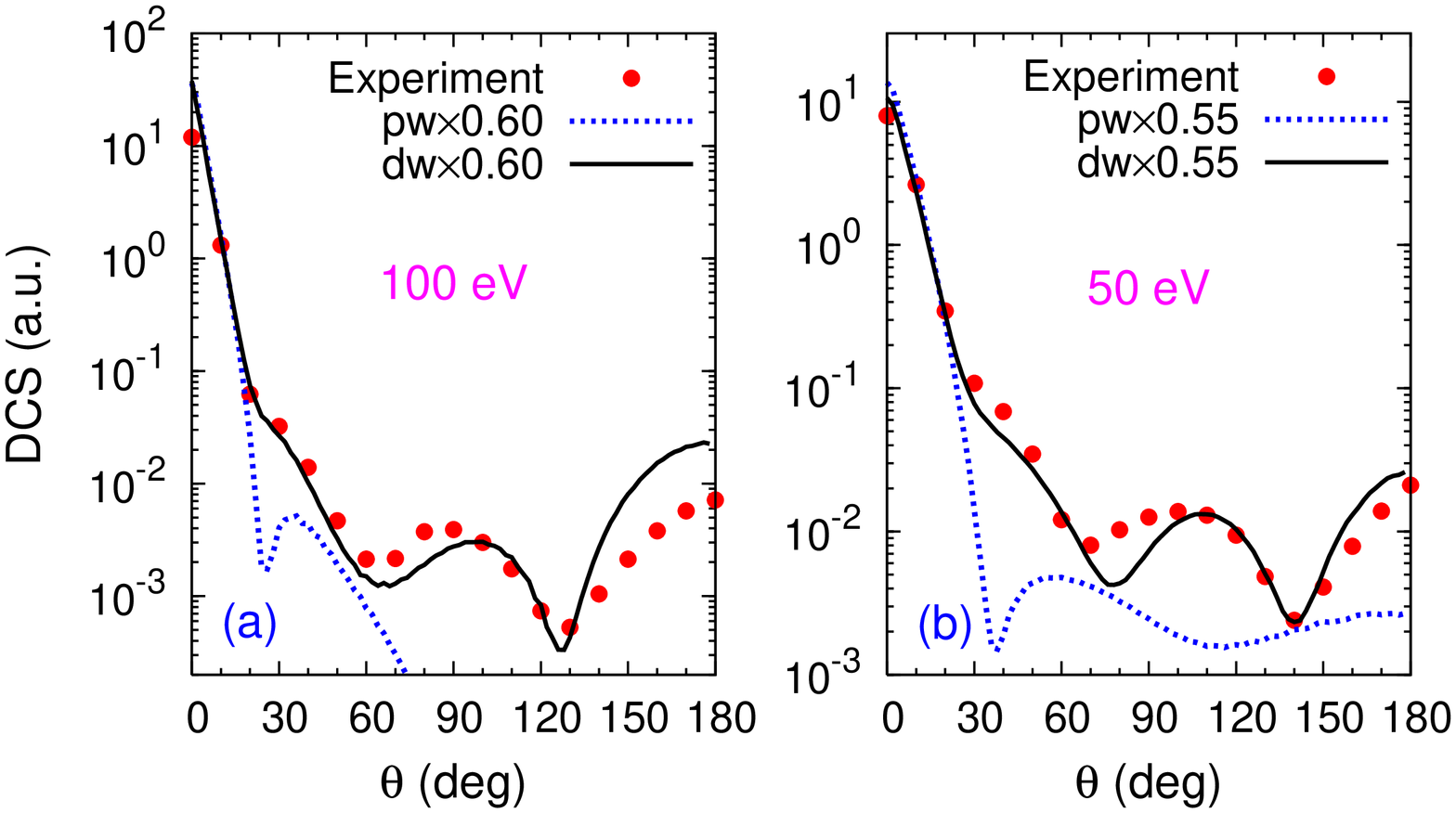}}}}
\mbox{\rotatebox{270}{\myscaleboxa{
\includegraphics{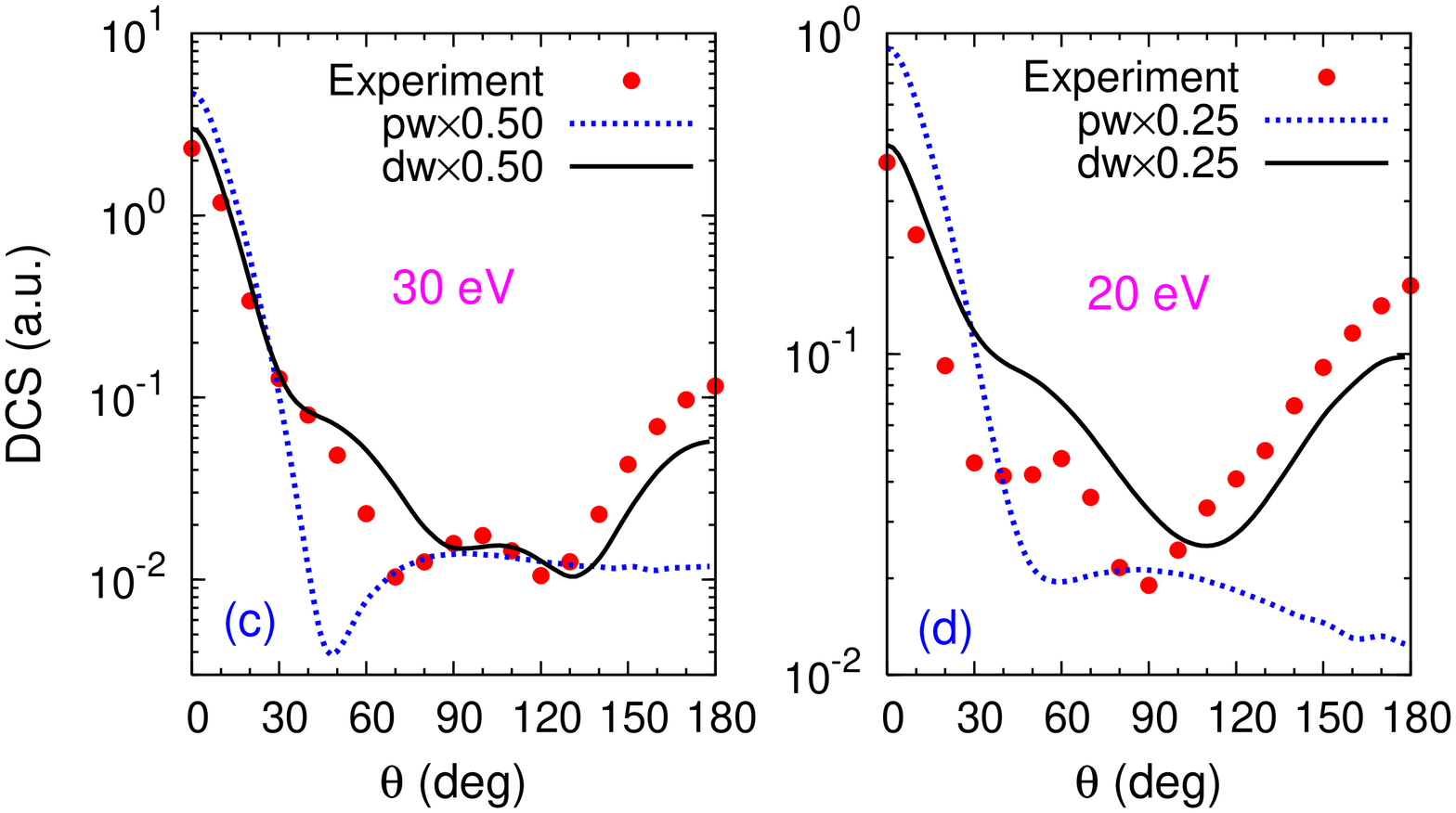}}}}
\caption{DCS for the excitation of the $3p^54s$ configuration of Ar
by electron impact at incident energies of (a) 100 eV, (b) 50 eV,
(c) 30 eV and (d) 20 eV. The absolute experimental measurements are
from Chutjian and Cartwright~\cite{Chutjain}.}
\end{figure}

\begin{figure}
\centering \mbox{\rotatebox{270}{\myscaleboxa{
\includegraphics{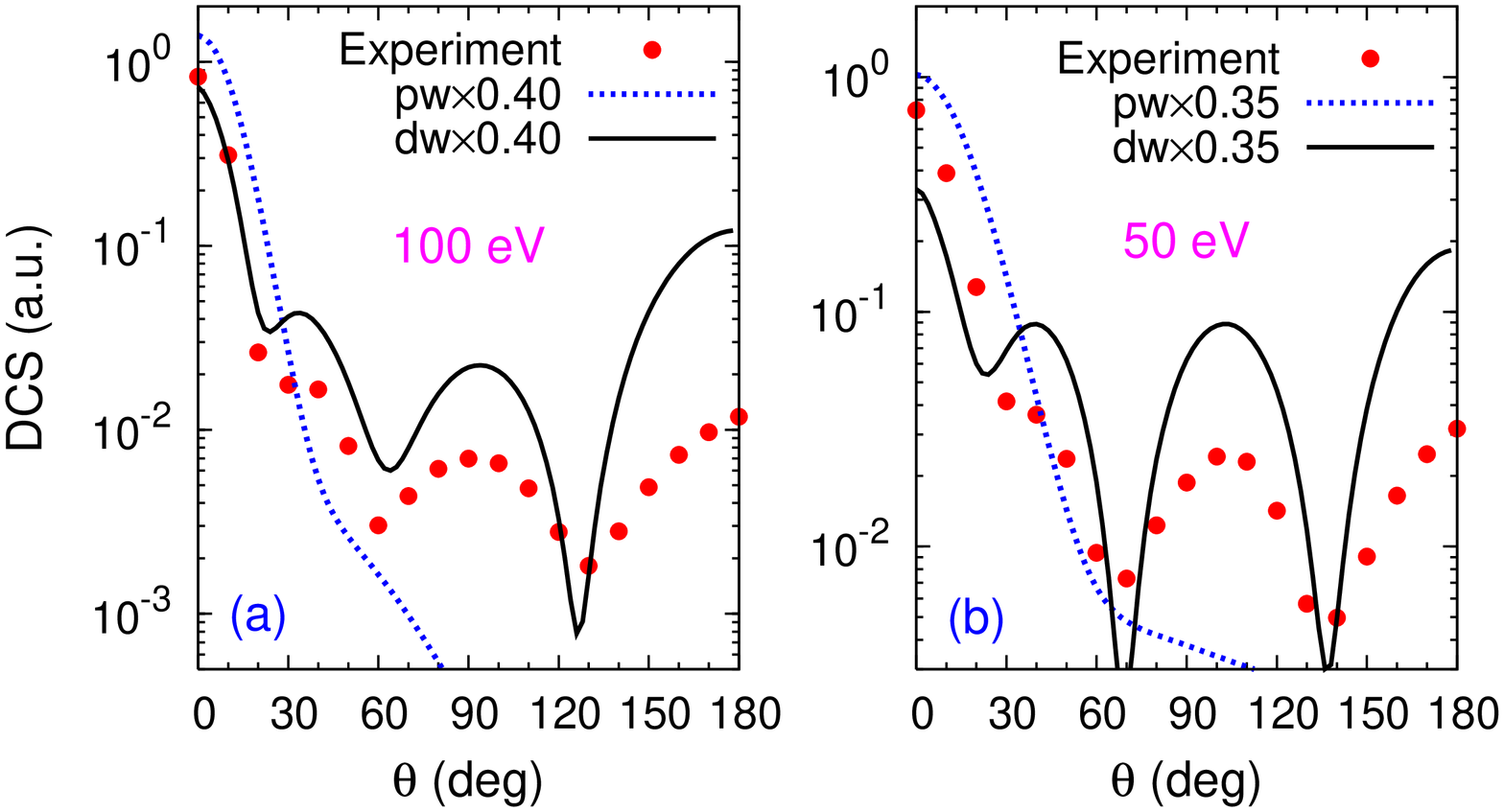}}}}
\mbox{\rotatebox{270}{\myscaleboxa{
\includegraphics{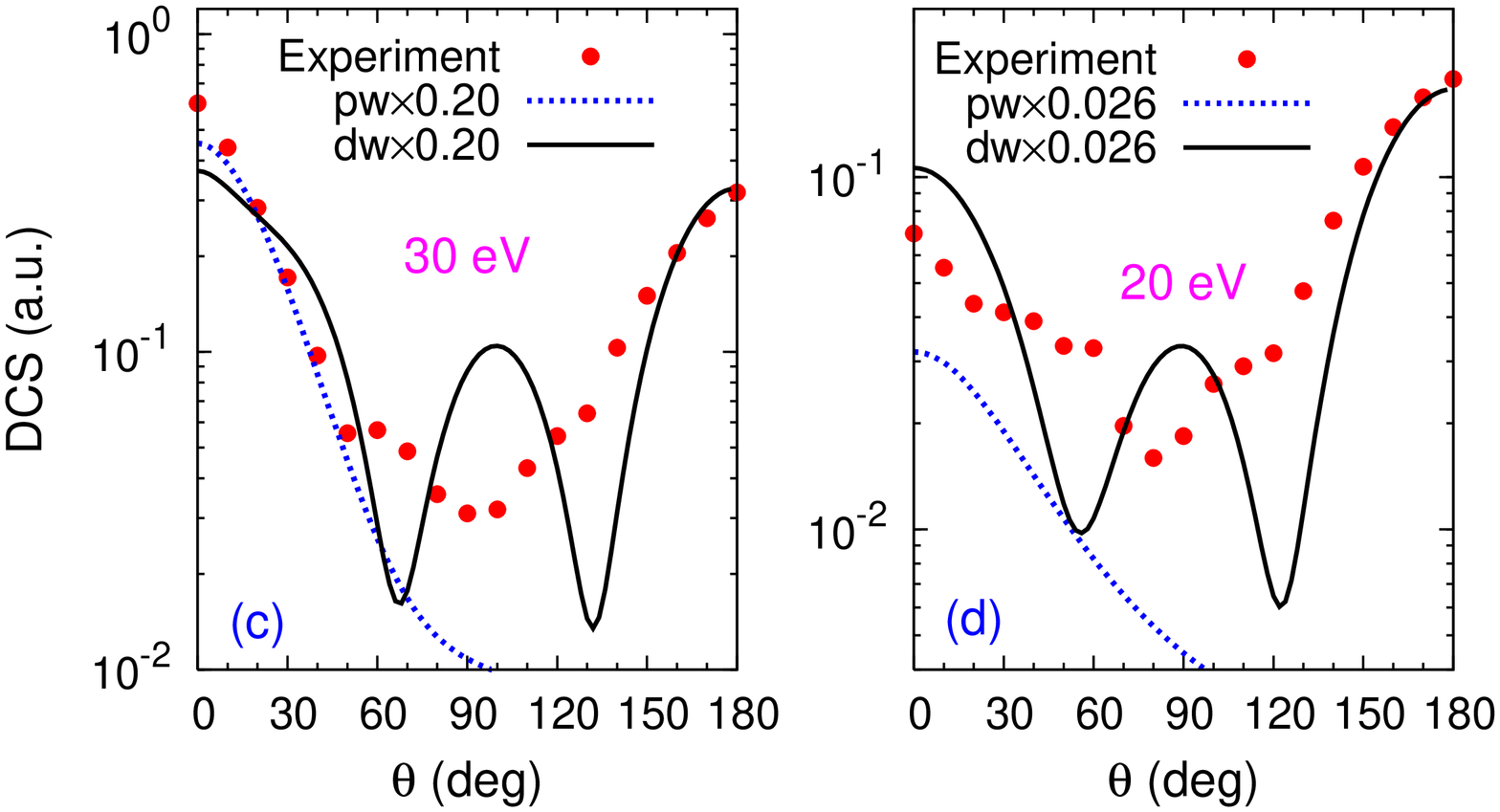}}}}
\caption{Same as Fig.~5 but for the $3p^54p$ configuration at
incident energies of (a) 100 eV, (b) 50 eV, (c) 30 eV, and (d) 20
eV.}
\end{figure}

In Figs.~5 and 6, we show similar comparison for the excitation of
$3p^54s$ and $3p^54p$ configurations of Ar. The experimental
measurements were performed by Chutjian and
Cartwright~\cite{Chutjain}. Compared to the excitation of Ne, the
DCS of Ar have more structures. For example, for $3p^54s$  at 100 eV
and 50 eV, as shown in Figs.~5(a,b), in addition to the rapid slope
change around 25$^{\circ}$, extra minima were observed in experiment
which are reproduced by the DWBA. For $3p^54p$ at 100 eV and 50 eV,
as shown in Figs.~6(a,b), DWBA predicts triple minima in the DCS.
They were observed in experiment as well despite that the
backscattering is overestimated by the DWBA. This might indicate
that the distorting potential used in the calculations needs to be
improved. For lower incident energies of 30 eV and 20 eV, the
agreement between DWBA and experiment for
  $3p^54p$ of Ar can not be regarded as satisfactory.
However, the main feature can still be predicted by the DWBA. It
should be noted that, for the excitation of $3p^54p$ of Ar at 20 eV
(see Fig.~6(d)), the yield from DWBA exceeds the experimental value by
an even greater margin than the PWBA result. For this case, the DWBA
predicts the DCS which is about 40 times higher than experiment.

\begin{figure}
\centering \mbox{\rotatebox{270}{\myscaleboxb{
\includegraphics{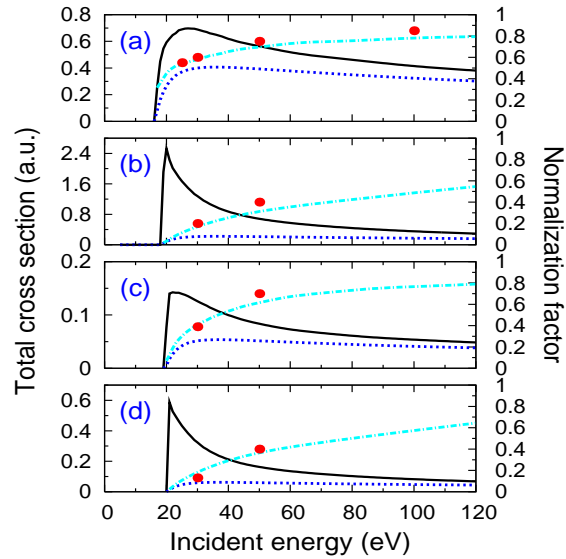}}}}
\caption{Total cross sections (left vertical axis) and normalization
factors of DWBA (right vertical axis) for electron impact excitation
of Ne from $2p^6$ to (a) $2p^53s$, (b) $2p^53p$, (c) $2p^54s$, and
(d) $2p^54p$. Solid curve, total cross sections of DWBA; Dotted
curve, total cross sections calculated using the empirical formula
of Tong \emph{et al}.~\cite{Tong03pra}; Chain curve, scaling factor
$\mathcal{C}(E_i)$; Solid circles, normalization factors used in
Figs.~1-4 for DWBA to obtain the best overall agreement with
experiment.}
\end{figure}

\begin{figure}
\centering \mbox{\rotatebox{270}{\myscaleboxb{
\includegraphics{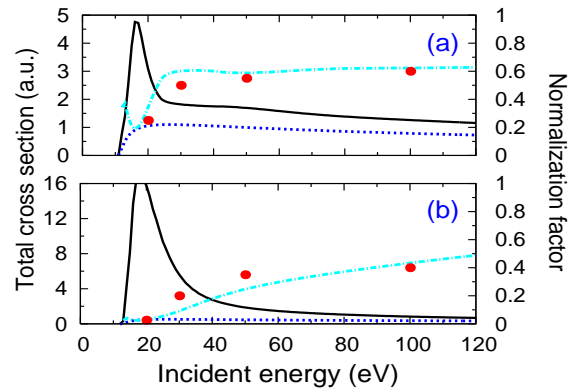}}}}
\caption{Same as Fig.~7 but for Ar from $3p^6$ to (a) $3p^54s$ and
(b) $3p^54p$. The solid circles are the normalization factors used
in Figs.~5 and 6 for DWBA to obtain the best overall agreement with
experiment.}
\end{figure}

The empirical formula, Eq.~(\ref{xm1}), has already been used to
calculate the total ionization yield of Ar in NSDI as a function of
the peak intensity for a linearly polarized laser pulse by Micheau
\emph{et al}.~\cite{Sam-NSDI}.

To obtain the scaling factor $\mathcal{C}(E_i)$ for electron-Ne
excitation, we calculate the TCS's using DWBA and those using
Eq.~(\ref{xm1}) for all the four configurations considered here at
incident energies from threshold up to 500 eV. These TCS's for
incident energies below 120 eV are plotted in Fig.~7 referring to
the left vertical axis. It can be seen that the difference in the
magnitude of TCS between DWBA and Tong \emph{et al} increases with
decreasing incident energy. This difference is indicated by the the
scaling factor $\mathcal{C}(E_i)$ which is also plotted in Fig.~7,
referring to the right vertical axis. To see the accuracy of the
scaling factor, the normalization factors used in Figs.~1-4 for DWBA
to obtain the best overall agreement with the experimental DCS's are
displayed for comparison. One can see that all the normalization
factors used in Figs.~1-4 agree well with those predicted by the
scaling factor $\mathcal{C}(E_i)$. In Fig.~8, similar comparisons
for Ar are shown. The good agreement of $\mathcal{C}(E_i)$ with the
normalization factors used in Figs.~5 and 6 confirms again the
validity of the calibration method.

In conclusion, we proposed a method to calibrate the DCS from DWBA
for electron impact excitation of atoms at low energies. The method
will be applied to simulate the correlated electron momentum spectra
for NSDI of atoms in a strong laser field, in which electron impact
excitation of the ions is involved. This work paves the way for
theoretical study, based on the QRS model, on NSDI of atoms in
strong laser pulse.

\section*{Acknowledgment}

This work was supported in part by Chemical Sciences, Geosciences
and Biosciences Division, Office of Basic Energy Sciences, Office of
Science, US Department of Energy. Y Liang was supported with
financial aids from China Scholarship Council, China Education
Ministry under Grant No. 2009-1590, and Education Department in
Liaoning Province of China under Grant No. 2009A305. The work of
D.H.M. was supported by the National Science Foundation under Grant
No. PHY-0757749.

\end{document}